\documentstyle[12pt,epsfig]{article}
\begin{document}
\topskip 2cm
\begin{titlepage}
\rightline{ \large{ \bf DFUB 98-5} }
\rightline{ \large{ \bf TP-USL/98/4} }
\rightline{ \large{ \bf May 1998} }
\begin{center}
{\large\bf The Master Differential Equations for the 2-loop Sunrise Selfmass 
Amplitudes. } \\
\vspace{2.5cm}
\begin{center}
{\large {\bf
M.~Caffo$^{ab}$, 
H.~Czy{\.z}\ $^{c \star}$,
S.~Laporta$^{b}$} and   
{\bf E.~Remiddi$^{ba}$ \\ } }
\end{center}

\begin{itemize}
\item[$^a$]
             {\sl INFN, Sezione di Bologna, I-40126 Bologna, Italy }
\item[$^b$] 
             {\sl Dipartimento di Fisica, Universit\`a di Bologna, 
             I-40126 Bologna, Italy }
\item[$^c$] 
             {\sl Institute of Physics, University of Silesia, 
             PL-40007 Katowice, Poland }

\end{itemize}
\end{center}

\noindent
e-mail: {\tt caffo@bo.infn.it \\ 
\hspace*{1.3cm} czyz@usctoux1.cto.us.edu.pl \\ 
\hspace*{1.3cm} laporta@bo.infn.it \\ 
\hspace*{1.3cm} remiddi@bo.infn.it \\ } 
\vspace{.5cm}
\begin{center}
\begin{abstract}
The master differential equations in the external square momentum \( p^2 \) 
for the master integrals of the two-loop sunrise graph, in 
\( n \)-continuous dimensions and for arbitrary values of the internal 
masses, are derived. The equations are then used for working out the 
values at \( p^2 = 0 \) and the expansions in \( p^2 \) at \( p^2 =0 \), in 
\( (n-4) \) at \( n\to4 \) limit and in \( 1/p^2 \) for large values of 
\( p^2 \). 
\end{abstract}
\end{center}
\scriptsize{ \noindent ------------------------------- 
} 
\vfill

\footnoterule
\noindent
$^{\star}${\footnotesize 
     Partly supported by the Polish Committee for Scientific Research
     under grant no 2P03B17708. } 
\end{titlepage}
\pagestyle{plain} \pagenumbering{arabic} 
\def\Li2{\hbox{Li}_2} 
\def\LLL{L(m_1^2,m_2^2,m_3^2)} 
\def\a{\alpha} 
\def\app{{\left(\frac{\alpha}{\pi}\right)}} 
\newcommand{\Eq}[1]{Eq.(\ref{#1})} 
\newcommand{\labbel}[1]{\label{#1}} 
\newcommand{\cita}[1]{\cite{#1}} 
\newcommand{\dnk}[1]{ \frac{d^nk_{#1}}{(2\pi)^{n-2}} } 
\newcommand{\e}{{\mathrm{e}}} 
\newcommand{\verso}[1]{ {\; \buildrel {n \to #1} \over{\longrightarrow}}\; } 
\section{Introduction.}           \par 
In a previous paper \cita{ER}, it was shown that a linear 
system of first order differential equations in the external Mandelstam 
variables, to be referred to from now on as master equations, can in 
general be written for the master integrals of any Feynman graph. In 
this paper the master equations are written for the 2-loop sunrise 
self-mass graph amplitudes in \( n \) continuous dimensions and for 
arbitrary values of the masses as a function of the square momentum transfer 
\( p^2 \). All momenta are Euclidean, so that \( p^2 > 0 \) corresponds 
to spacelike \( p_\mu \). 
The master equations can be used for obtaining valuable information 
on the properties of the master integrals, for working out virtually any 
desired expansion, for obtaining closed analytic results (when possible) 
and, in any case, for the actual numerical calculation of the amplitudes. 
\par 
In the various sections of this paper we will show 
\begin{itemize} 
\item how to use the integration by part identities \cita{ChetTka} 
      , \cita{Tarasov} for writing the master equations;
\item how to obtain the (analytic) values of the master integrals at 
      \( p^2 = 0 \), in the \( n\to4 \) limit; 
\item how to expand the master integrals in \( p^2 \) at \( p^2 = 0 \); 
\item how to expand the master integrals in \( (n-4) \) for arbitrary 
      \( p^2 \); 
\item how to solve analytically the master equations when two masses 
      are equal to zero; 
\item how to expand the master integrals in \( 1/p^2 \) for large values 
      of $p^2$ and arbitrary values of the masses. 
\end{itemize} 
\par 
The appendix contains a few ready-to-use recurrence formulae, obtained 
by solving the integration by parts identities for the considered 
graph \cita{Tarasov}. \par 
All the algebra needed to generate the identities, 
the equations and the considered expansions has been processed by means 
of the computer program FORM \cita{FORM} by J. Vermaseren. 
\par 
While we deal with the 2 loop case only, it should be apparent from the 
discussion that some suitable extension of the results may apply also to 
more general cases. 
\section{The master integrals and their differential equations.} \par 
The 2-loop sunrise self-mass graph is shown in Fig.1. \\ 

\begin{figure}[!h]
\epsfbox[0 20 140 140]{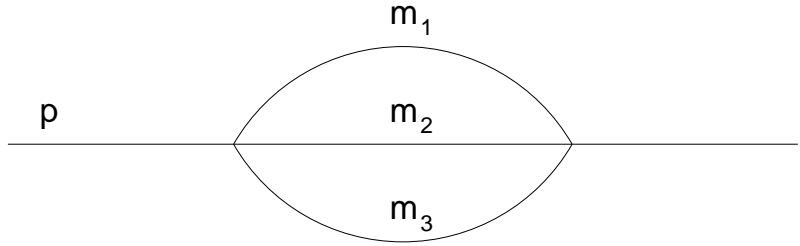}
\caption{The 2-loop sunrise self-mass graph.} 
\end{figure}

The related amplitudes, in \( n \) continuous dimensions and fully Euclidean 
variables, can be written as 
\begin{eqnarray} 
  A(n,m_1^2,m_2^2,m_3^2,p^2,-\alpha_1,-\alpha_2,-\alpha_3,\beta_1,\beta_2) 
      &=& \nonumber \\  
      && {\kern-200pt} \int \dnk{1} \int \dnk{2} \; 
      \frac{ (p\cdot k_1)^{\beta_1} (p\cdot k_2)^{\beta_2} } 
           { (k_1^2+m_1^2)^{\alpha_1} (k_2^2+m_2^2)^{\alpha_2} 
             ( (p-k_1-k_2)^2+m_3^2 )^{\alpha_3} } \ . 
\labbel{1} \end{eqnarray} 
\newcommand{\F}[1]{F_{#1}(n,m_1^2,m_2^2,m_3^2,p^2)} 
\newcommand{\Fn}[2]{F_{#1}^{(#2)}(m_1^2,m_2^2,m_3^2,p^2)} 
\newcommand{\Fp}[2]{F_{{#1},{#2}}(n,m_1^2,m_2^2,m_3^2)} 
\newcommand{\Fpn}[3]{F_{{#1},{#2}}^{(#3)}(m_1^2,m_2^2,m_3^2)} 
\newcommand{\Pol}[2]{P_{#1,#2}(m_1^2,m_2^2,m_3^2,p^2)} 
\newcommand{\Qol}[2]{Q_{#1,#2}(m_1^2,m_2^2,m_3^2,p^2)} 
\newcommand{\Qbl}[2]{\bar Q_{#1,#2}(m_1^2,m_2^2,m_3^2,p^2)} 
\newcommand{\D}{D(m_1^2,m_2^2,m_3^2,p^2)} 
\newcommand{\R}[1]{R^{#1}(m_1^2,m_2^2,m_3^2)} 
It is by now well known that, as a consequence of the integration by parts 
identities \cita{ChetTka}, all the above amplitudes can be expressed in 
terms of four master integrals \cita{Tarasov}, which can be taken as 
\begin{eqnarray} 
  \F{0} &=& A(n,m_1^2,m_2^2,m_3^2,p^2,-1,-1,-1,0,0) \ , \nonumber \\ 
  \F{1} &=& A(n,m_1^2,m_2^2,m_3^2,p^2,-2,-1,-1,0,0) \ , \nonumber \\ 
  \F{2} &=& A(n,m_1^2,m_2^2,m_3^2,p^2,-1,-2,-1,0,0) \ , \nonumber \\ 
  \F{3} &=& A(n,m_1^2,m_2^2,m_3^2,p^2,-1,-1,-2,0,0) \ . 
\labbel{2} \end{eqnarray} 
It follows from the definitions that the last three master integrals 
are the mass derivatives of the first, {\sl i.e.} for \( i = 1, 2, 3 \) 
\begin{equation}
  \F{i} = - \frac{\partial}{\partial m_i^2} \F{0} \ . 
\labbel{3} \end{equation}
The simplest way for deriving the required differential equations for 
self-mass amplitudes, which depend on a single Mandelstam variable only, 
consists in writing for them the familiar scaling equation (which can be 
obtained, for instance, by scaling in the definition \Eq{1} all the 
integration momenta by a same factor \( \lambda \) and then differentiating 
with respect to \( \lambda \)). \\ 
For \( \F{0} \) that procedure gives 
\begin{equation} 
  \left(     p^2 \frac{\partial}{\partial p^2} 
         + m_1^2 \frac{\partial}{\partial m_1^2} 
         + m_2^2 \frac{\partial}{\partial m_2^2} 
         + m_3^2 \frac{\partial}{\partial m_3^2} 
         - (n-3) \right) \F{0} = 0 \ , 
\labbel{4} \end{equation} 
{\sl i.e.}, on account of Eq.(\ref{3}), 
\begin{eqnarray} 
  p^2 \frac{\partial}{\partial p^2} \F{0} &=& (n-3) \F{0} + m_1^2 \F{1} 
                                           \nonumber \\ 
                                          &+& m_2^2 \F{2} + m_3^2 \F{3} \ . 
\labbel{5} \end{eqnarray} 
A similar equation holds for each of $\F{i}$, $i = 1, 2, 3$ 
\begin{equation} 
  \left(     p^2 \frac{\partial}{\partial p^2} 
         + m_1^2 \frac{\partial}{\partial m_1^2} 
         + m_2^2 \frac{\partial}{\partial m_2^2} 
         + m_3^2 \frac{\partial}{\partial m_3^2} 
         - (n-4) \right) \F{i} = 0 \ . 
\labbel{6} \end{equation} 
When carrying out the mass derivatives, new integrals appear, like for 
instance those corresponding to \( \alpha_1 = 3 \), 
\( \alpha_2 = \alpha_3 = 1 \) and \( \beta_1 = \beta_2 = 0 \) in 
the notation of Eq.(\ref{1}). All those integrals, according to the 
formulas of the Appendix, can be expressed in terms of the master 
integrals. The result can be written as 
\\ \vbox{ \begin{eqnarray} 
  p^2 \frac{\partial}{\partial p^2} \F{i} &=& \frac{n-4}{2} \F{i} 
      \nonumber \\ 
  && {\kern-165pt} + \frac{n-3}{\D} \left[ \; \phantom{+} \Pol{i}{0} 
                          \frac{3n-8}{2} \F{0} \right. \nonumber \\ 
  && {\kern-50pt} + \Pol{i}{1} \F{1} \nonumber \\ 
  && {\kern-50pt} + \Pol{i}{2} \F{2} \nonumber \\ 
  && {\kern-50pt} \left. + \Pol{i}{3} \F{3} \; \right] \nonumber \\ 
  && {\kern-165pt} + \frac{(n-2)^2}{\D} \left[ \; \phantom{+} 
                    \Qol{i}{1} T(n,m_2^2) T(n,m_3^2) \right. \nonumber \\ 
  && {\kern-50pt} + \Qol{i}{2} T(n,m_1^2) T(n,m_3^2) \nonumber \\ 
  && {\kern-50pt} \left. 
                  + \Qol{i}{3} T(n,m_1^2) T(n,m_2^2) \; \right] \ , 
\labbel{7} \end{eqnarray} } 
where \( \D \), the \( \Pol{i}{j} \) and the \( \Qol{i}{j} \) are 
polynomials in \( m_1^2, m_2^2, m_3^2 \) and \( p^2 \), to be discussed in 
a moment, while \( T(n,m^2) \) is defined by 
\begin{equation} 
  T(n,m^2) = \int\dnk{ } \frac{1}{k^2+m^2} = \frac{m^{n-2}}{(n-2)(n-4)}C(n)\ , 
\labbel{16} \end{equation} 
\( C(n) \) being the following function of \( n \) 
\begin{equation} 
 C(n) = \left(2 \sqrt{\pi} \right)^{(4-n)} \Gamma\left(3-\frac{n}{2}\right) \ . 
\labbel{16a} \end{equation} 
The expansion of \Eq{16a} in \( n \) around \( n=4 \) is 
\begin{eqnarray}
  C(n) &=& 1 - (n-4) \left[ \log(2 \sqrt{\pi}) - \frac{1}{2} \gamma_E 
          \right] \nonumber\\ 
        &+&(n-4)^2 \frac{1}{2} \left[ \log^2(2 \sqrt{\pi}) 
    +\frac{1}{4} \left(\frac{\pi^2}{6} + \gamma_E^2 \right)
    -\gamma_E \log(2 \sqrt{\pi}) \right] \nonumber\\ 
    &+& {\cal O}\left( (n-4)^3 \right) \ , 
\labbel{16b} \end{eqnarray}
where \( \gamma_E \) is the Euler-Mascheroni constant. 
The expansion, which is given only to make easier the comparisons with other 
results in the literature, is however never needed in practice, despite 
the frequent presence of \( 1/(n-4) \) factors, as in \Eq{16}. 
Indeed, there is always an overall factor \( C(n) \) for each loop; 
in the calculation of any physical (and therefore finite) quantity, 
the singularities, which are of the form \( C^2(n)/(n-4)^2 \) etc., 
cancel out, and in the remaining finite part it is sufficient to use the 
value at \( n=4 \), which is simply 
\begin{equation} C(4) = 1 \ . \labbel{17} \end{equation} 
\par 
The explicit expression of the polynomial \( \D \) is 
\begin{eqnarray} 
  \D &=& (p_2)^4\ +\ 4(m_1^2+m_2^2+m_3^2)(p^2)^3 \nonumber \\ 
     && {\kern-100pt} + 2\left( 3m_1^4+3m_2^4+3m_3^4+2m_1^2m_2^2 
                      +2m_1^2m_3^2+2m_2^2m_3^2 \right)(p^2)^2 \nonumber \\ 
     && {\kern-100pt} + 4(m_1^6+m_2^6+m_3^6-m_1^4m_2^2-m_1^4m_3^2-m_1^2m_2^4 
     -m_1^2m_3^4-m_2^4m_3^2-m_2^2m_3^4                         \nonumber \\ 
     && {\kern-100pt} \phantom{+ 4(} + 10m_1^2m_2^2m_3^2) p^2 + \R{4} 
                                                              \nonumber \\ 
     && {\kern-100pt} = \left(p^2+(m_1+m_2+m_3)^2\right) \ 
                        \left(p^2+(m_1+m_2-m_3)^2\right) \nonumber \\ 
     && {\kern-100pt} \phantom{=} \cdot \left(p^2+(m_1-m_2+m_3)^2\right) \ 
                        \left(p^2+(m_1-m_2-m_3)^2\right) \ ,
\labbel{8} \end{eqnarray} 
where 
\begin{eqnarray} 
  \R{2} &=& m_1^4+m_2^4+m_3^4-2m_1^2m_2^2-2m_1^2m_3^2-2m_2^2m_3^2 
                                                              \nonumber \\ 
  &=& \phantom{\cdot}(m_1+m_2+m_3)(m_1+m_2-m_3)               \nonumber \\ 
  &\phantom{=}& \cdot(m_1-m_2+m_3)(m_1-m_2-m_3) \ . 
\labbel{8a} \end{eqnarray} 
The polynomial $P_{i,j}$, for \( i = 1 \) are given by 
\begin{eqnarray} 
  \Pol{1}{0} &=& - (p^2)^3\ -\ (3m_1^2+m_2^2+m_3^2) (p^2)^2 \nonumber \\ 
  && {\kern-100pt} - (3m_1^4-m_2^4-m_3^4-2m_1^2m_2^2-2m_1^2m_3^2 
                                         +10m_2^2m_3^2) p^2 \nonumber \\ 
  && {\kern-100pt} - (m_1^2-m_2^2-m_3^2)\R{2} \ , \nonumber \\
  \Pol{2}{0} &=& P_{1,0}(m_2^2,m_1^2,m_3^2,p^2) \ , \nonumber \\ 
  \Pol{3}{0} &=& P_{1,0}(m_3^2,m_2^2,m_1^2,p^2) \ , 
\labbel{9} \end{eqnarray} 
\vbox{ \begin{eqnarray} 
  \Pol{1}{1} &=& - (2m_1^2+m_2^2+m_3^2) (p^2)^3 \nonumber \\ 
  && {\kern-100pt} - (6m_1^4+3m_2^4+3m_3^4+3m_1^2m_2^2+3m_1^2m_3^2 
                            +2m_2^2m_3^2) (p^2)^2 \nonumber \\ 
  && {\kern-100pt} - (6m_1^6+3m_2^6+3m_3^6 -5m_1^4m_2^2-5m_1^4m_3^2 
          -4m_1^2m_2^4-4m_1^2m_3^4 \nonumber \\ 
  && {\kern-100pt} \phantom{- (} 
          -3m_2^4m_3^2-3m_2^2m_3^4 +40m_1^2m_2^2m_3^2) p^2 \nonumber \\ 
  && {\kern-100pt} - (2m_1^4+m_2^4+m_3^4-3m_1^2m_2^2-3m_1^2m_3^2 
                                        -2m_2^2m_3^2)\R{2} \ , \nonumber \\
  \Pol{2}{2} &=& P_{1,1}(m_2^2,m_1^2,m_3^2,p^2) \ , \nonumber \\
  \Pol{3}{3} &=& P_{1,1}(m_3^2,m_2^2,m_1^2,p^2) \ , 
\labbel{10} \end{eqnarray} } 
\vbox{ \begin{eqnarray} 
  \Pol{1}{2} &=& m_2^2\left[ -3(p^2)^3 - (7m_1^2+5m_2^2-3m_3^2)(p^2)^2 
                      \right. \nonumber \\ 
  && {\kern-100pt} 
     -(5m_1^4+m_2^4-7m_3^4-6m_1^2m_2^2+2m_1^2m_3^2+14m_2^2m_3^2) 
       p^2 \nonumber \\ 
  && {\kern-100pt} \left. - (m_1^2-m_2^2-m_3^2)\R{2} \right] \ , 
\labbel{11} \end{eqnarray} } 
\begin{eqnarray} 
  \Pol{1}{3} &=& P_{1,2}(m_1^2,m_3^2,m_2^2,p^2) \ , \nonumber \\
  \Pol{2}{1} &=& P_{1,2}(m_2^2,m_1^2,m_3^2,p^2) \ , \nonumber \\
  \Pol{2}{3} &=& P_{1,2}(m_2^2,m_3^2,m_1^2,p^2) \ , \nonumber \\
  \Pol{3}{1} &=& P_{1,2}(m_3^2,m_1^2,m_2^2,p^2) \ , \nonumber \\
  \Pol{3}{2} &=& P_{1,2}(m_3^2,m_2^2,m_1^2,p^2) \ ; 
\labbel{12} \end{eqnarray} 
similarly, the $Q_{i,j}$ are given by 
\begin{eqnarray} 
  \Qol{1}{1} &=& \Qol{2}{2} \ = \ \Qol{3}{3} \ = \nonumber \\
&=& - \frac{1}{2}\left[ 3(p^2)^2 + 2(m_1^2+m_2^2+m_3^2) p^2 -\R{2} \right] \ , 
\labbel{13} \end{eqnarray} 
\begin{eqnarray} 
  \Qol{1}{2} &=& \frac{1}{4m_1^2}\left[ (p^2)^3 + (m_1^2+m_2^2+3m_3^2)(p^2)^2 
                              \right.      \nonumber \\ 
  && {\kern-100pt} 
  -(m_1^4+m_2^4-3m_3^4-10m_1^2m_2^2+2m_1^2m_3^2+2m_2^2m_3^2) p^2 \nonumber \\ 
  && {\kern-100pt} \left. - (m_1^2+m_2^2-m_3^2)\R{2}  \right] \ , 
\labbel{14} \end{eqnarray} 
\begin{eqnarray} 
  \Qol{1}{3} &=& Q_{1,2}(m_1^2,m_3^2,m_2^2,p^2) \ , \nonumber \\
  \Qol{2}{1} &=& Q_{1,2}(m_2^2,m_1^2,m_3^2,p^2) \ , \nonumber \\
  \Qol{2}{3} &=& Q_{1,2}(m_2^2,m_3^2,m_1^2,p^2) \ , \nonumber \\
  \Qol{3}{1} &=& Q_{1,2}(m_3^2,m_1^2,m_2^2,p^2) \ , \nonumber \\
  \Qol{3}{2} &=& Q_{1,2}(m_3^2,m_2^2,m_1^2,p^2) \ .
\labbel{15} \end{eqnarray} 
The values corresponding to the other indices can be obtained by a suitable 
permutation of the masses. 
\par 
The above polynomials satisfy the relations 
\begin{eqnarray} 
  && m_2^2 m_3^2 \ \Qol{1}{1} \ = \ - \frac{1}{4} (m_1^2 - m_2^2 - m_3^2 ) 
                              \Pol{1}{0} \nonumber \\ 
  && {\kern+15pt} + \frac{1}{8}\left( \Pol{1}{1} - \Pol{1}{2} 
                            - \Pol{1}{3} \right) \ , \nonumber \\ 
  && m_1^2 m_3^2 \ \Qol{1}{2} \ = \  \frac{1}{4} (m_1^2 - m_2^2 + m_3^2 ) 
                              \Pol{1}{0} \nonumber \\ 
  && {\kern+15pt} - \frac{1}{8}\left( \Pol{1}{1} - \Pol{1}{2} 
                            + \Pol{1}{3} \right) \ , \nonumber \\ 
  && m_1^2 m_2^2 \ \Qol{1}{3} \ = \  \frac{1}{4} (m_1^2 + m_2^2 - m_3^2 ) 
                              \Pol{1}{0} \nonumber \\ 
  && {\kern+15pt} - \frac{1}{8}\left( \Pol{1}{1} + \Pol{1}{2} 
                            - \Pol{1}{3} \right) \ , 
\labbel{15a} \end{eqnarray} 
and 
\begin{eqnarray} 
  && \D = - ( p^2 - m_1^2 - m_2^2 - m_3^2 ) \Pol{1}{0} \nonumber \\ 
  && {\kern+15pt} - \Pol{1}{1} - \Pol{1}{2} - \Pol{1}{3} \ . 
\labbel{15b} \end{eqnarray} 
\par 
The equations Eq.s(\ref{5},\ref{7}) are the master differential 
equations referred to in the title of this paper. 
They are long, but their structure is simple: 
they express the derivatives with respect to \( p^2 \) of the four master 
integrals as a linear combination of the master integrals themselves and of 
known non-homogenous terms, the coefficients being ratios of known 
polynomials in the masses and \( p^2 \). 
The known terms are just products of the 1-loop, 1-denominator vacuum 
amplitude \( T(n,m^2) \) of Eq.(\ref{16}), the simplest integral in the 
game which fixes the overall normalization. \par 
All the coefficients of the linear combinations are equal to a simple 
factor, depending on the dimension \( n \) times a ratio of polynomials 
in \( p^2 \) and the masses. There is a unique denominator, the polynomial 
\( \D \) of \Eq{8}; note that its zeros in \( p^2 \) correspond to 
the threshold and the three pseudo-thresholds of the considered 2-loop 
sunrise graph. 
\par 
It should be apparent, from the derivation, that master equations of 
similar structure, 
({\it i.e.} equations whose coefficients are ratios of polynomials) 
can be established for the master integrals of virtually any Feynman 
graph, for any number of loops; the size of the polynomials is expected 
to grow quickly with the number of the denominators and of the loops, but they 
are in any case finite polynomials, which can be conveniently dealt with 
by means of an algebraic program. Quite in general, the non-homogeneous 
terms correspond to amplitudes of related Feynman graphs in which one of 
the propagators is missing; from this point of view, the $l$-loop sunrise 
amplitudes are the simplest $l$-loop amplitudes, in the sense that in 
their equations the non-homogeneous terms are just products of $l-1$ 
1-denominator vacuum amplitudes, Eq.(\ref{16}). Sunrise amplitudes are 
in turn present as non-homogeneous terms in the equations for the amplitudes 
of graphs having one more propagator and so on. 
\par 
As it will be shown in the rest of the paper, the master 
equations contain virtually all the information needed for 
studying the analytic properties of the master integrals and for obtaining 
any expansion of interest; furthermore, after some minor work (the 
expansion in \( (n-4) \) at \( n=4 \)), the differential equations can be 
used for the actual numerical evaluation of the master integrals, 
without explicit reference to the original integral representation 
\Eq{1}, once the values of the integrals are known at some initial 
value of \( p^2 \). Both the expansion in \( (n-4) \) and the 
evaluation of the initial values (at \( p^2=0 \)), can be carried out 
by exploiting the information contained in the master equations (which 
hold identically in \( n \) and are regular at \( p^2 = 0 \) despite the 
appearance of an apparent kinematical singularity at that value of 
\( p^2 \)). 
\section{The value at \( p^2=0 \).}           \par 
From the definitions Eq.s(\ref{1},{2}), it is obvious that all the master 
integrals as well as all their derivatives are regular at \( p^2 = 0 \) 
(which in the considered Euclidean region corresponds to \( p_\mu = 0 \)). 
If we introduce the vacuum graph of Fig.2 \\ 
\pagebreak

\begin{figure}[!h]
\epsfbox[0 20 140 80]{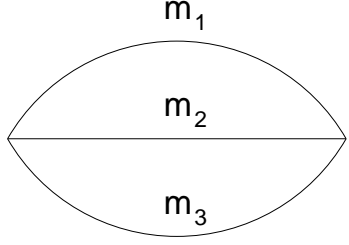} 
\caption{The 2-loop sunrise vacuum graph.} 
\end{figure}

and the corresponding amplitude 
\begin{equation} 
  V(n,m_1^2,m_2^2,m_3^2) = \int \dnk{1} \int \dnk{2} \; 
      \frac{1}{ (k_1^2+m_1^2)(k_2^2+m_2^2)( (k_1+k_2)^2+m_3^2 ) } \ , 
\labbel{18} \end{equation} 
the \( p^2=0 \) values of the master integrals defined in Eq.s(\ref{2}) 
can then be written as 
\begin{eqnarray} 
  F_0(n,m_1^2,m_2^2,m_3^2,0)   &=& V(n,m_1^2,m_2^2,m_3^2) \ , 
  \labbel{19a} \\ 
  F_i(n,m_1^2,m_2^2,m_3^2,0) &=& - \frac{\partial}{\partial m_i^2} 
 V(n,m_1^2,m_2^2,m_3^2) \ . 
\labbel{19} \end{eqnarray} 
Let us now consider the differential equations Eq.(\ref{5}) and Eq.(\ref{7}) 
for, say, $i = 3$ at \( p^2=0 \). 
Due to the regularity of the derivatives, the left hand sides of both 
equations vanish; therefore, the right hand sides must also vanish at 
\( p^2=0 \). 
It turns out that both equations involve the same combination of 
\( F_1(n,m_1^2,m_2^2,m_3^2,0) \) and \( F_2(n,m_1^2,m_2^2,m_3^2,0) \), namely 
\( m_1^2F_1(n,m_1^2,m_2^2,m_3^2,0) + m_2^2F_2(n,m_1^2,m_2^2,m_3^2,0) \); 
by eliminating that combination, one is left with a single equation 
involving only \( F_0(n,m_1^2,m_2^2,m_3^2,0) \) and 
\( F_3(n,m_1^2,m_2^2,m_3^2,0) \), which is its derivative with respect to 
\( m_3^2 \), according to \Eq{19} ; 
in terms of the above vacuum amplitude the equation reads 
\begin{eqnarray} 
 \R{2} \frac{\partial}{\partial m_3^2}V(n,m_1^2,m_2^2,m_3^2) &=& \nonumber \\ 
 && {\kern-160pt} \phantom{+ } (n-3)( m_3^2-m_1^2-m_2^2 )
 V(n,m_1^2,m_2^2,m_3^2)    \nonumber \\ 
  && {\kern-160pt} + (n-2)\frac{m_3^2-m_1^2+m_2^2}{2m_3^2}T(n,m_1^2)T(n,m_3^2)
                                              \nonumber \\ 
  && {\kern-160pt} + (n-2)\frac{m_3^2+m_1^2-m_2^2}{2m_3^2}T(n,m_2^2)T(n,m_3^2)
                                              \nonumber \\ 
  && {\kern-160pt} - (n-2)T(n,m_1^2)T(n,m_2^2) \ , 
\labbel{20} \end{eqnarray} 
where \( \R{2} \) is the same as in \Eq{8a}. 
\Eq{20} is the master equation for the 2-loop vacuum amplitude of Fig.(2) 
-- this time in one of the masses (the differential 
equations in the other masses are immediately obtained by a suitable 
permutation of the masses themselves). It is also easy to see that the 
same equations can be obtained by writing down directly the integration 
by part identities for \( V(n,m_1^2,m_2^2,m_3^2) \), Eq.(\ref{18}) and 
the related amplitudes. \par 
Eq.(\ref{20}) can be used for evaluating $V(n,m_1^2,m_2^2,m_3^2) $ once its 
``initial" value corresponding to a particular value of \( m_3^2 \) is given. 
Note that the equation itself can provide such a value. 
Indeed, the vacuum amplitude and all its derivatives are regular for 
non-vanishing values of the masses; 
by considering the particular value \( m_3=m_1+m_2 \), 
the factor \( \R{2} \) on the {\it l.h.s.} of Eq.(\ref{20}) vanish, so 
that the {\it r.h.s.} must also vanish; that amounts to the relation 
\begin{eqnarray} 
  V(n,m_1^2,m_2^2,(m_1+m_2)^2) &&= \frac{n-2}{2(n-3)} 
   \left[ \frac{T(n,m_1^2)}{m_1} \frac{T(n,m_2^2)}{m_2} \right. \nonumber \\ 
 && \left. - \left( \frac{T(n,m_1^2)}{m_1} +\frac{T(n,m_2^2)}{m_2} \right) 
     \frac{T(n,(m_1+m_2)^2)}{(m_1+m_2)} \right] \ . 
\labbel{21} \end{eqnarray} 
\Eq{21} can be taken as the ``initial condition" which determines 
completely the solution of \Eq{20}. Therefore, the differential equation 
\Eq{20}, a first order equation in a single unknown function, can be solved 
in quadrature using the initial value given by \Eq{21}. \par 
For the explicit calculations it is convenient to expand the amplitude 
\( V(n,m_1^2,m_2^2,m_3^2) \) around \( n=4 \), where it is expected to 
develop polar singularities in \( (n-4) \). More exactly, \Eq{16} shows 
that \( T(n,m_i^2) \) has a single pole \( (n-4) \), so that the non 
homogenous part of \Eq{20} and the initial value 
\( V(n,m_1^2,m_2^2,(m_1+m_2)^2) \), \Eq{21}, both develop double poles; 
it follows that also \( V(n,m_1^2,m_2^2,m_3^2) \) can have at most double 
poles. It can therefore be expanded as 
\begin{eqnarray} 
  V(n,m_1^2,m_2^2,m_3^2) = && C^2(n) \Biggl\{ 
           \frac{1}{(n-4)^2} V^{(-2)}(m_1^2,m_2^2,m_3^2)
         + \frac{1}{(n-4)}   V^{(-1)}(m_1^2,m_2^2,m_3^2) \nonumber \\ 
      && + V^{(0)}(m_1^2,m_2^2,m_3^2) +{\cal O} (n-4)  \Biggr\} \ , 
\labbel{22} \end{eqnarray} 
where \( C(n) \) is given by \Eq{16} and \( C^2(n) \) has been 
treated as an overall factor according to the remarks preceding \Eq{17}. 
Correspondingly, \Eq{20} splits into the following set of equations 
for the residua and the finite part at \( n=4 \) 
\begin{eqnarray} 
  \R{2} && \frac{\partial}{\partial m_3^2} V^{(-2)}(m_1^2,m_2^2,m_3^2) = 
 (m_3^2-m_1^2-m_2^2) V^{(-2)}(m_1^2,m_2^2,m_3^2) 
\nonumber \\ && 
   -\frac{1}{4} \left[ m_1^2 (m_1^2-m_3^2)+m_2^2 (m_2^2-m_3^2) \right] \ ,
\labbel{23} \end{eqnarray} 
\begin{eqnarray} 
  \R{2} && \frac{\partial}{\partial m_3^2} V^{(-1)}(m_1^2,m_2^2,m_3^2) = 
   (m_3^2-m_1^2-m_2^2) V^{(-1)}(m_1^2,m_2^2,m_3^2) 
\nonumber \\ && 
 + (m_3^2-m_1^2-m_2^2) V^{(-2)}(m_1^2,m_2^2,m_3^2) 
\nonumber \\ && 
 +\frac{1}{8} \Bigl\{ m_1^2 (m_1^2-m_3^2)+m_2^2 (m_2^2-m_3^2) 
\nonumber \\ && 
 -\left[ m_1^2 \log(m_1^2) +m_2^2 \log(m_2^2) \right] (m_1^2 +m_2^2 -m_3^2 ) 
\nonumber \\ && 
 - \log(m_3^2) \left[ m_1^2 (m_1^2-m_2^2-m_3^2) + m_2^2 (-m_1^2+m_2^2-m_3^2)
\right]            \Bigr\} \ ,
\labbel{24} \end{eqnarray} 
\begin{eqnarray} 
 \R{2} && \frac{\partial}{\partial m_3^2} V^{(0)}(m_1^2,m_2^2,m_3^2) = 
    (m_3^2-m_1^2-m_2^2)  V^{(0)}(m_1^2,m_2^2,m_3^2) 
\nonumber \\ && 
  + (m_3^2-m_1^2-m_2^2)  V^{(-1)}(m_1^2,m_2^2,m_3^2) 
\nonumber \\ && {\kern-100pt} 
 +\frac{1}{16} \biggl\{ -m_1^2 (m_1^2-m_3^2)-m_2^2 (m_2^2-m_3^2) 
 - 2 m_1^2 m_2^2 \log(m_1^2) \log(m_2^2) 
\nonumber \\ && {\kern-100pt} 
 +\biggl[ m_1^2 \log(m_1^2) \left( 1-\frac{1}{2} \log(m_1^2) \right) 
        + m_2^2 \log(m_2^2) \left( 1-\frac{1}{2} \log(m_2^2) \right) \biggr] 
(m_1^2 +m_2^2 -m_3^2) 
\nonumber \\ && {\kern-100pt} 
 + \log(m_3^2) \left( 1-\frac{1}{2} \log(m_3^2) \right)  
   \left[ m_1^2 (m_1^2-m_2^2-m_3^2) + m_2^2 (-m_1^2+m_2^2-m_3^2) \right]
\nonumber \\ && {\kern-100pt} 
 - \log(m_3^2) \Bigl[ m_1^2 (m_1^2-m_2^2-m_3^2)  \log(m_1^2) 
                     +m_2^2 (-m_1^2+m_2^2-m_3^2) \log(m_2^2) \Bigr] 
                   \biggr\} \ . 
\labbel{25} \end{eqnarray} 
Similarly, in the notation of \Eq{22} the coefficients of the expansion 
of \Eq{21} are 
\begin{eqnarray} 
 V^{(-2)}(m_1^2,m_2^2,(m_1+m_2)^2) &=& 
          - \frac{1}{4}(m_1 m_2+m_1^2+m_2^2) \ , \nonumber \\ 
 V^{(-1)}(m_1^2,m_2^2,(m_1+m_2)^2) &=& 
            \frac{3}{8}(m_1 m_2+m_1^2+m_2^2)  \nonumber \\ && 
 {\kern-100pt} - \frac{1}{4} \left[ m_1^2 \log(m_1) +m_2^2 \log(m_2) 
             +(m_1+m_2)^2 \log(m_1+m_2) \right] \ ,
\nonumber \\ 
 V^{(0)}(m_1^2,m_2^2,(m_1+m_2)^2) &=&  
          - \frac{7}{16}(m_1 m_2+m_1^2+m_2^2) \nonumber \\ && 
 {\kern-100pt} +\frac{3}{8} \left[ m_1^2 \log(m_1) +m_2^2 \log(m_2) 
             +(m_1+m_2)^2 \log(m_1+m_2) \right] 
\nonumber \\ && 
 {\kern-100pt} -\frac{1}{8} \left[ m_1 \log(m_1) - m_2 \log(m_2) \right]^2
               -\frac{1}{8} (m_1+m_2)^2 \log^2(m_1+m_2) 
\nonumber \\ && 
{\kern-100pt}  -\frac{1}{4} \left[ m_1 \log(m_1) + m_2 \log(m_2) \right]
                                  (m_1+m_2) \log(m_1+m_2) \ . 
 \labbel{25c} \end{eqnarray} 
When solved recursively, starting from \( j = -2 \), the equations 
(\ref{23},\ref{24},\ref{25}) are all of the form 
\begin{eqnarray} 
  \R{2} \frac{\partial}{\partial m_3^2} V^{(j)}(m_1^2,m_2^2,m_3^2) &=& 
 (m_3^2-m_1^2-m_2^2) V^{(j)}(m_1^2,m_2^2,m_3^2) 
\nonumber \\ && 
 + g^{(j)}(m_1^2,m_2^2,m_3^2)  \ ,
\labbel{25a} \end{eqnarray} 
where the $g^{(j)}(m_1^2,m_2^2,m_3^2), j = -2, -1, 0$, to be identified 
by comparison with the original equations (\ref{23},\ref{24},\ref{25}), can 
be considered as known. \par 
\Eq{25a} can in principle be solved by the quadrature formula
\begin{eqnarray} 
  V^{(j)}(m_1^2,m_2^2,m_3^2) &=& \R{} \Biggl\{ 
 \frac{V^{(j)}(m_1^2,m_2^2,z)}{R(m_1^2,m_2^2,z)}  
\nonumber \\ 
&+& \int_{z}^{m_3^2} \frac{dx}{R^3(m_1^2,m_2^2,x)}
 g^{(j)}(m_1^2,m_2^2,x)  \Biggr\}   \ ,
\labbel{25b} \end{eqnarray} 
where $z$ is any initial point. Our initial point is \( (m_1+m_2)^2 \); as 
for that value of \( z \) the two terms in the {\it r.h.s.} diverge, 
it is convenient to modify the formula by carrying out an integration 
by parts and then to take the $z = (m_1+m_2)^2$ limit. When that is done 
\Eq{25b} becomes 
\begin{eqnarray} 
  V^{(j)}(m_1^2,m_2^2,m_3^2) =&& {\kern-20pt} \frac{1}{4 m_1^2 m_2^2} 
  \Biggl\{  \R{} {\kern-15pt} \int\limits_{(m_1+m_2)^2}^{m_3^2} 
           {\kern-15pt} dx \ \frac{x-m_1^2-m_2^2}{R(m_1^2,m_2^2,x)} \ 
           \frac{\partial}{\partial x} g^{(j)}(m_1^2,m_2^2,x)  \nonumber \\ 
&& + (m_1^2+m_2^2-m_3^2) g^{(j)}(m_1^2,m_2^2,m_3^2)  \Biggr\}   \ ;
\labbel{25bb} \end{eqnarray} 
that last formula, as can be easily verified, satisfies the differential 
equations (\ref{23},\ref{24},\ref{25}), with the initial conditions 
\Eq{25c}, implied by the regularity at $m_3 = m_1+m_2$. \par 
The integration can be actually carried out by using the change of variable 
\begin{equation} 
  x = m_1^2 + m_2^2 + m_1 m_2 \frac{t^2 + 1}{t} 
 = \frac{m_1 m_2}{t} \left( t +\frac{m_1}{m_2} \right) 
                     \left( t +\frac{m_2}{m_1} \right) \ ,
\labbel{26} \end{equation} 
which can be inverted into
\begin{eqnarray} 
  t = t (x;m_1^2,m_2^2) 
    &&= \frac{\sqrt{x - (m_1 - m_2)^2} +\sqrt{x - (m_1 + m_2)^2}}
             {\sqrt{x - (m_1 - m_2)^2} -\sqrt{x - (m_1 + m_2)^2}}
\nonumber \\ 
    &&= \frac{1}{2 m_1 m_2} \left( x-m_1^2-m_2^2 + R(x,m_1^2,m_2^2) 
                            \right) \ . 
\labbel{27} \end{eqnarray} 
The solutions of the equations (\ref{23},\ref{24},\ref{25}) with the initial 
conditions given by the equations (\ref{25c}) are then respectively 
\begin{equation} 
  V^{(-2)}(m_1^2,m_2^2,m_3^2) = - \frac{1}{8} (m_1^2+m_2^2+m_3^2) \ ,
\labbel{29} \end{equation} 
\begin{eqnarray} 
  V^{(-1)}(m_1^2,m_2^2,m_3^2) && = 
       \frac{1}{8} \Biggl\{ \frac{3}{2} (m_1^2+m_2^2+m_3^2) 
\nonumber \\ && 
     - \left[ m_1^2 \log(m_1^2) +m_2^2 \log(m_2^2) +m_3^2 \log(m_3^2) \right]
                        \Biggr\} \ , 
\labbel{30} \end{eqnarray} 
\begin{eqnarray} 
 V^{( 0)}(m_1^2,m_2^2,m_3^2) \quad =
       \frac{1}{16} && \biggl\{ \R{} \LLL -\frac{7}{2} (m_1^2+m_2^2+m_3^2) 
\nonumber \\ && 
     + 3 \left[ m_1^2 \log(m_1^2) +m_2^2 \log(m_2^2) +m_3^2 \log(m_3^2) \right]
\nonumber \\ && 
 -\frac{1}{2} 
 \left[ m_1^2 \log^2(m_1^2) +m_2^2 \log^2(m_2^2) +m_3^2 \log^2(m_3^2) \right]
\nonumber \\ && 
 +\frac{1}{2} \Bigl[ (-m_1^2 -m_2^2 +m_3^2) \log(m_1^2) \log(m_2^2) 
\nonumber \\ && 
                   + (-m_1^2 +m_2^2 -m_3^2) \log(m_1^2) \log(m_3^2) 
\nonumber \\ && 
                   + ( m_1^2 -m_2^2 -m_3^2) \log(m_2^2) \log(m_3^2) \Bigr]
                        \biggr\} \ , 
\labbel{31} \end{eqnarray} 
where
\begin{eqnarray}  
 &&\LLL =\Li2(- t_3 m_2/m_1) +\Li2(- t_3 m_1/m_2) +\zeta(2) 
                             +\frac{1}{2} \log^2(t_3)
\nonumber \\ && 
 +\frac{1}{2} \left[ \log(t_3+m_2/m_1) -\log(t_3+m_1/m_2) 
 +\frac{3}{4} \log(m_1^2/m_2^2) \right] \log(m_1^2/m_2^2) \ , 
\labbel{31a} \end{eqnarray} 
with \( t_3=t(m_3^2;m_1^2,m_2^2) \), see \Eq{27} . 
It is to be noted that Eq.s(\ref{29}-\ref{31}) are expected to be 
symmetrical in the exchange of $m_1^2$, $m_2^2$, $m_3^2$; the symmetry, 
which is manifest for most terms, holds also for $\LLL$, \Eq{31a}, 
even if in this case the symmetry is not manifest. 
Note also that \Eq{31a} remains valid in its present form for any 
real positive value of $m_1$, $m_2$ and $m_3$, provided that \( t_3 \) 
is given by the second of Eq.s(\ref{27}). In the limiting case 
\( m_1,m_2 \to 0 \), finally, independently of the order of the limits 
one finds 
\begin{equation} 
 V^{( 0)}(0,0,m^2) \quad = - \frac{m^2}{16} \left[ \zeta(2) + \frac{7}{2} 
                           - 3\log(m^2) + \log^2(m^2) \right] \ . 
\labbel{31b} \end{equation} \par 
Analytical values of $V(n,m_1^2,m_2^2,m_3^2)$ were already presented in 
\cita{FJJ}, where a differential equations method based on mass derivatives 
\cita{Kotikov} was used, and in \cita{DT}, where the results were obtained 
by using a Mellin-Barnes representation and its subsequent expansion in 
\( n \to 4 \) limit. Our results agree numerically with those presented 
in \cita{FJJ} in terms of Lobachevskiy's function and a related one; 
to check that it agrees also with the analytic formula of \cita{DT} 
it is necessary to use the following relation between dilogarithms of 
different arguments  
\begin{eqnarray} 
&&   \Li2(- t_2 m_1 /m_3) + \Li2(- t_1 m_2 /m_3) 
    - \log(- t_2 m_1 /m_3)    \log(- t_1 m_2 /m_3) = 
\nonumber \\ &&
 \Li2(- t_3 m_1 /m_2)  + \Li2(- t_3 m_2 /m_1)   + 2 \zeta(2)
 +\frac{1}{2} \log^2(t_3) 
 -\frac{1}{2} \log(m_1^2/m_3^2) \log(m_2^2/m_3^2) 
\nonumber \\ &&
 +\frac{1}{2} \left[ \log(t_3+m_2/m_1) - \log(t_3+m_1/m_2) \right] 
 \log(m_1^2/m_2^2) 
 +\frac{3}{8} \log^2(m_1^2/m_2^2) \ , 
\labbel{33} \end{eqnarray} 
valid for $ m_3 \geq m_1+m_2 $, with \( t_1=t(m_1^2;m_2^2,m_3^2) \)
and \( t_2=t(m_2^2;m_1^2,m_3^2) \), whose validity can be established 
by standard methods. \par 
\section{The expansion at \( p^2 \simeq 0 \).}           \par 
As \( p^2 = 0 \) is a regular point for the 4 functions \( \F{i} \), 
\( i=0,1,2,3 \), they can be expanded in powers of \( p^2 \) around that 
point 
\begin{equation} 
  F_i(n,m_1^2,m_2^2,m_3^2,p^2) 
   = \sum \limits_{k=0}^{\infty} \Fp{i}{k} (p^2)^k \ . 
\labbel{34} \end{equation} 
The very first term, according to \Eq{19a} is 
\begin{equation} 
 \Fp{0}{0}= F_0(n,m_1^2,m_2^2,m_3^2,0) = V(n,m_1^2,m_2^2,m_3^2) \ , 
\labbel{34.1} \end{equation} 
where the vacuum amplitude $V(n,m_1^2,m_2^2,m_3^2)$ 
discussed and evaluated in the previous Section, \Eq{22} and 
Eq.s(\ref{29}-\ref{31}). Once 
$V(n,m_1^2,m_2^2,m_3^2)$ is known, \Eq{19} and \Eq{20} can be 
combined to give, if \( (i,j,k) \) is any permutation of the indices 
\( (1,2,3) \) 
\begin{eqnarray} 
  \Fp{i}{0} &=& F_i(n,m_1^2,m_2^2,m_3^2,0) \nonumber \\ 
            &=& - \frac{1}{\R{2}} \left[ 
     \phantom{\frac{m_i}{m_j}} {\kern-15pt} 
    (n-3)( m_i^2-m_j^2-m_k^2 ) V(n,m_1^2,m_2^2,m_3^2)  \right.  \nonumber \\ 
  && + (n-2)\frac{m_i^2-m_j^2+m_k^2}{2m_i^2}T(n,m_j^2)T(n,m_i^2)
                                              \nonumber \\ 
  && + (n-2)\frac{m_i^2+m_j^2-m_k^2}{2m_i^2}T(n,m_k^2)T(n,m_i^2)
                                              \nonumber \\ 
  && \left. - (n-2)T(n,m_j^2)T(n,m_k^2) \phantom{\frac{m_i}{m_j}} \right] \ , 
\labbel{20a} \end{eqnarray} 
showing that the three quantities \( \Fp{i}{0} \), \( i=1,2,3, \) are all 
expressed algebraically in terms of \( V(n,m_1^2,m_2^2,m_3^2) \) only. \par 
When the expansion \Eq{34} is inserted in the Eq.s (\ref{4}) and (\ref{7}), 
by imposing the equality of the coefficients of the various powers of 
\( p^2 \) one obtains a set of equations for the quantities 
\( \Fp{i}{k} \). For \( k = 0 \) one recovers of course the equations 
discussed and solved in the previous Section; for \( k > 0 \) one gets a set 
of linear algebraic equations for the \( \Fp{i}{k} \), which can be solved 
recursively in terms of the \( \Fp{i}{k-1} \). Therefore everything can 
be expressed in terms of the \( \Fp{i}{0} \), which in turn depend, as 
already remarked, on the single vacuum amplitude 
\( V(n,m_1^2,m_2^2,m_3^2) \); as a consequence, all the \( \Fp{i}{k} \) 
can be expressed in terms of \( V(n,m_1^2,m_2^2,m_3^2) \) only. \par 
The first iteration of the just sketched procedure gives 
\begin{eqnarray} 
  \Fp{0}{1} &=& \frac{n-3}{n\R{4}} \Biggl[ (n-4) \left( 
                 m_1^4(m_2^2+m_3^2-m_1^2)   \right. \nonumber \\ 
     && {\kern-110pt} \left. + \ 
            m_2^4(m_1^2-m_2^2+m_3^2) + m_3^4(m_1^2+m_2^2-m_3^2) \right) 
               - 2(n-8)m_1^2m_2^2m_3^2 \Biggr] V(n,m_1^2,m_2^2,m_3^2) 
                                            \nonumber \\ 
  && {\kern-110pt} + \frac{n-2}{n\R{4}}\left[ \ \ 
        (n-4)(m_3^2+m_1^2-m_2^2)(m_3^2-m_1^2+m_2^2) \right. \nonumber \\ 
  && {\kern+70pt} \left. 
            + 2m_3^2(m_1^2+m_2^2-m_3^2) \right] T(n,m_1^2)T(n,m_2^2) 
\nonumber \\ 
  && {\kern-110pt} + \frac{n-2}{n\R{4}}\left[ \ \ 
        (n-4)(m_2^2+m_1^2-m_3^2)(m_2^2-m_1^2+m_3^2) \right. \nonumber \\ 
  && {\kern+70pt} \left. 
            + 2m_2^2(m_1^2-m_2^2+m_3^2) \right] T(n,m_1^2)T(n,m_3^2) 
\nonumber \\ 
  && {\kern-110pt} + \frac{n-2}{n\R{4}}\left[ \ \ 
        (n-4)(m_1^2+m_2^2-m_3^2)(m_1^2-m_2^2+m_3^2) \right. \nonumber \\ 
  && {\kern+70pt} \left. 
            + 2m_1^2(m_2^2-m_1^2+m_3^2) \right] T(n,m_2^2)T(n,m_3^2) \ . 
\labbel{34a} \end{eqnarray}
As already done for \( V(n,m_1^2,m_2^2,m_3^2) \) in \Eq{22}, the 
\( \Fp{i}{k} \) can be in turn expanded in $(n-4)$ as 
\begin{eqnarray} 
 \Fp{i}{k} = && C^2(n) \Biggl\{ \frac{1}{(n-4)^2} \Fpn{i}{k}{-2}
                + \frac{1}{(n-4)}   \Fpn{i}{k}{-1} \nonumber \\
             && + \Fpn{i}{k}{0} +{\cal O} (n-4) \Biggr\} \ , 
\labbel{35} \end{eqnarray} 
where \( C(n) \) is given as usual by \Eq{16a} and \Eq{17}. From \Eq{34a} 
and Eq.s(\ref{29}-\ref{31}) one obtains 
\begin{equation} 
   \Fpn{0}{1}{-2} = 0 \ ,
\labbel{36} \end{equation} 
\begin{equation} 
   \Fpn{0}{1}{-1} = \frac{1}{32} \ , 
\labbel{37} \end{equation} 
\begin{eqnarray} 
&& \Fpn{0}{1}{0} \quad =
       \frac{1}{32} \biggl\{ \frac{4 m_1^2 m_2^2 m_3^2}{\R{3}} \LLL 
                                - \frac{3}{4}
\nonumber \\ && 
 +\frac{1}{\R{2}} \Bigl[  m_1^2 ( m_1^2 -m_2^2 -m_3^2) \log(m_1^2) 
\nonumber \\ && 
                        + m_2^2 (-m_1^2 +m_2^2 -m_3^2) \log(m_2^2) 
                        + m_3^2 (-m_1^2 -m_2^2 +m_3^2) \log(m_3^2) 
                  \Bigr] \biggr\} \ . 
\labbel{38} \end{eqnarray} 
This result is in numerical agreement with the first term in the 
$p^2/m_3^2$ expansion in \cita{Berends}. 
It is straightforward to obtain, when required, any other term of the 
double expansion in \( p^2 \) and \( (n-4) \). 
\section{The expansion in $ (n-4) $ for arbitrary $p^2$.}  \par 
Quite in general, by using the same argument leading to the expansion 
of \Eq{22}, the functions \( \F{i}, i=0,1,2,3, \) are expected to develop 
up to a double pole in \( (n-4) \) and can be expanded, also for 
arbitrary \( p^2 \), in the by now customary form 
\begin{eqnarray} 
 && \F{i} = C^2(n) \Biggl\{  \frac{1}{(n-4)^2} \Fn{i}{-2} 
            \nonumber \\ 
 && {\kern+50pt} + \frac{1}{(n-4)}   \Fn{i}{-1} 
                  + \Fn{i}{0} + {\cal O} (n-4) \Biggr\} \ . 
\labbel{39a} \end{eqnarray} 
When all the terms of Eq.s(\ref{5}, \ref{7}) are consistently expanded in 
\( (n-4) \), one obtains a set of equations for the residua and the 
finite part in \( (n-4) \). Those equations can then 
be solved recursively, starting 
from the double pole. The equations corresponding to the poles in 
\( (n-4) \) will not be written explicitly here for the sake of brevity; 
it is however easy to verify that the solutions of those equations, 
corresponding to the coefficients of the poles in \( (n-4) \) 
of \( \F{0} \) are 
\begin{eqnarray} 
 \Fn{0}{-2} &=& -\frac{1}{8} (m_1^2+m_2^2+m_3^2) \ ,
\nonumber \\
 \Fn{0}{-1} &=&  \frac{1}{8} \Biggl\{ \frac{p^2}{4} 
                +\frac{3}{2} (m_1^2+m_2^2+m_3^2) 
\nonumber \\ && 
     - \left[ m_1^2 \log(m_1^2) +m_2^2 \log(m_2^2) +m_3^2 \log(m_3^2) \right]
                        \Biggr\} \ , 
\labbel{39b} \end{eqnarray} 
from which the corresponding coefficients \( \Fn{i}{k} \), for 
\( i=1,2,3 \) and \( k=-2,-1 \) are immediately obtained by using 
the Eq.s(\ref{3}),  
\begin{eqnarray} 
 \Fn{i}{-2} &=& \frac{1}{8} \ ,
\nonumber \\
 \Fn{i}{-1} &=&  - \frac{1}{16} + \frac{1}{8} \log(m_i^2) \ . 
\labbel{39c} \end{eqnarray} 
The above functions, indeed, satisfy the differential equations and the 
initial condition at \( p^2=0 \) provided by Eq.s(\ref{29}-\ref{31}), 
therefore they are the exact solutions for arbitrary $p^2$. \par 
On account of the above results for the singular parts, 
from Eq.s(\ref{5}, \ref{7}) one obtains 
for the finite parts \( \Fn{i}{0} \) of the 4 amplitudes the equations 
\begin{eqnarray} 
       p^2 \frac{\partial}{\partial p^2} \Fn{0}{0} &=& 
               \Fn{0}{0} + \Fn{0}{-1} \nonumber \\ 
  && {\kern-150pt} + m_1^2 \Fn{1}{0} + m_2^2 \Fn{2}{0} + m_3^2 \Fn{3}{0} \ , 
\labbel{39_a} \end{eqnarray} 
and 
\\ \vbox{ \begin{eqnarray} 
 {\kern+25pt} && {\kern-25pt} 
    8   \D p^2 \frac{\partial}{\partial p^2} \Fn{i}{0} = \nonumber \\ 
 && \phantom{+}  4   \D \Fn{i}{-1} \nonumber \\ 
 && + \Pol{i}{0} \left[ 16   \Fn{0}{0} + 28   \Fn{0}{-1} \right. \nonumber \\ 
 && {\kern+228pt} \left. + 12   \Fn{0}{-2} \right] \nonumber \\ 
 && + 8 \Pol{i}{i} \left[ \Fn{i}{0} + \Fn{i}{-1} \right] \nonumber \\
 && + 8 \Pol{i}{j} \left[ \Fn{j}{0} + \Fn{j}{-1} \right] \nonumber \\ 
 && + 8 \Pol{i}{k} \left[ \Fn{k}{0} + \Fn{k}{-1} \right] \nonumber \\ 
 && + \Qol{i}{i} \ m_j^2 m_k^2 \left[ \log(m_j^2) + \log(m_k^2) \right]^2 
    \nonumber \\ 
 && + \Qol{i}{j} \ m_i^2 m_k^2 \left[ \log(m_i^2) + \log(m_k^2) \right]^2 
    \nonumber \\ 
 && + \Qol{i}{k} \ m_i^2 m_j^2 \left[ \log(m_i^2) + \log(m_j^2) \right]^2 \ , 
\labbel{39_b} \end{eqnarray} } 
where \( (i,j,k) \) is any permutation of \( (1,2,3) \) and  
\( \Pol{i}{j} \), \( \Qol{i}{j} \) are the polynomials of \Eq{7}. \par 
In the notation of \Eq{39a}, for \( j=-2,-1,0 \) 
\begin{equation} 
 F_0^{(j)}(m_1^2,m_2^2,m_3^2,0) = V^{(j)}(m_1^2,m_2^2,m_3^2) \ , 
\labbel{39aa} \end{equation} 
where the \( V^{(j)}(m_1^2,m_2^2,m_3^2) \) are given in 
Eq.s(\ref{29}-\ref{31}), so that \( F_0^{(0)}(m_1^2,m_2^2,m_3^2,0) \) 
is known; similarly, the \( F_i^{(0)}(m_1^2,m_2^2,m_3^2,0) \), 
\( i=1,2,3, \) can be obtained by using either \Eq{19} or \Eq{20a} 
(properly expanded in the \( n \to 4 \) limit). As a conclusion, 
\Eq{39_a} and \Eq{39_b}, together with the initial conditions at 
\( p^2=0 \) just discussed, are free by construction from the 
singularities of the $n \to 4$ limit and therefore ready to be used for 
the numerical evaluation of the finite parts of the sunrise master 
integrals. \par
For convenience of later use, we introduce one more quantity 
\( {\cal{F}}_0(m_1^2,m_2^2,m_3^2,p^2) \), defined by 
\begin{eqnarray} 
  {\cal{F}}_0(m_1^2,m_2^2,m_3^2,p^2) = \lim\limits_{n\to4} && {\kern-15pt} 
           \left[ {\kern+5pt} \F{0}  \right. \nonumber \\ 
                   && - \left. \Fp{0}{0} - p^2\Fp{0}{1} \right] \ , 
\labbel{39} \end{eqnarray} 
where \( \Fp{0}{0} \) and \( \Fp{0}{1} \) are given by \Eq{34.1} and 
\Eq{34a}. \par 
It is to be noted that the terms of \( \F{0} \) singular in \( (n-4) \) 
are at most linear 
in \( p^2 \), as shown in Eq.s(\ref{39b}), and therefore entirely 
encapsulated in the combination \( \Fp{0}{0} + p^2\Fp{0}{1} \); for that 
reason \( {\cal{F}}_0(m_1^2,m_2^2,m_3^2,p^2) \) is finite in 
the \( n \to 4 \) limit. \par 
From the very definition \Eq{39}, \( {\cal{F}}_0(m_1^2,m_2^2,m_3^2,p^2) \) 
has a double zero at \( p^2 = 0 \); 
therefore it satisfies the doubly subtracted dispersion relation 
\begin{equation} 
  {\cal{F}}_0(m_1^2,m_2^2,m_3^2,p^2)  = \left(p^2\right)^2 
  \int\limits_{(m_1+m_2+m_3)^2}^{\infty} \frac{d u}{u^2 (u+p^2)} 
  \ \frac{1}{\pi} \ \ {\mathrm{Im}} F_0(4,m_1^2,m_2^2,m_3^2,-u) \ , 
\labbel{40} \end{equation} 
which is obviously finite at \( n=4 \). \par 

\section{The analytic calculation for \( m_2 = m_3 = 0 \). } \par 
We discuss now the case in which all the masses but one are equal to zero; 
the remarkably simpler equations obtained in that limit can be solved 
analytically, providing useful information for the study of the large 
\( p^2 \) behaviour, even for arbitrary masses, which will be worked out 
in the next Section. \par 
When putting \( m_2 = m_3 = 0 \), and \( m_1 = m \) for simplicity, the 
system of the 4 equations \Eq{5} and \Eq{7} collapses 
into just two equations: 
\begin{eqnarray}
   p^2 \frac{\partial}{\partial p^2} F_0(n,m^2,0,0,p^2) &&{\kern-20pt}=
    m^2 F_1(n,m^2,0,0,p^2) + (n-3) F_0(n,m^2,0,0,p^2)   \nonumber \\
  p^2 \left(p^2 + m^2 \right) \frac{\partial}{\partial p^2} 
    F_1(n,m^2,0,0,p^2) &&{\kern-20pt}=
 - \frac{1}{2} \left(n-3  \right)\left(3n-8  \right) F_0(n,m^2,0,0,p^2)
    \nonumber \\ 
  &&{\kern-100pt}  +\left[ 2 p^2 (n-3) - (p^2+m^2) \left(\frac{3}{2}n
     -4 \right) \right] 
   F_1(n,m^2,0,0,p^2)   \ .
\labbel{199} \end{eqnarray}
The system so obtained is homogeneous, and as such it cannot provide 
information on the initial values; but the required initial conditions 
are still supplied by the proper \( m_2=m_3=0 \) limit of the results 
of the previous sections. \par 
To solve the system, we observe that it is of course equivalent to 
a second order differential equation in a single function. 
If \( F_0(n,m^2,0,0,p^2) \) is kept as the independent function, eliminating 
\(F_1(n,m^2,0,0,p^2)\) from Eq.s(\ref{199}) one obtains 
\begin{eqnarray}
&&{\kern-30pt}p^2(p^2+m^2) \frac{\partial^2}{\partial (p^2)^2}
  F_0(n,m^2,0,0,p^2) + \left(-2p^2(n-3)  + \frac{n}{2} (p^2+m^2)\right)
        \frac{\partial}{\partial p^2}  F_0(n,m^2,0,0,p^2)
\nonumber \\ 
&&+ \frac{(n-4)(n-3)}{2}  F_0(n,m^2,0,0,p^2) = 0 \ . 
\labbel{204} \end{eqnarray} 
By introducing \( {\cal{F}}_0(m^2,0,0,p^2) \) according to \Eq{39}, 
with the values of 
\( F_{0,0}(n,0,0,m^2) \) and \( F_{0,1}(n,0,0,m^2) \) corresponding 
to Eq.s(\ref{34.1}),(\ref{34a}) in the \( m_1=m_2=0 \) limit given by 
\Eq{22} and \Eq{31b}, one has 
\begin{eqnarray}
  F_0(n,m^2,0,0,p^2) = && C^2(n) \Biggl\{ - \frac{1}{(n-4)^2} \frac{m^2}{8} 
  + \frac{1}{(n-4)} \frac{1}{32}\left(p^2 
        + 6m^2 -4m^2 \log(m^2)\right) \nonumber\\ 
    &&{\kern-100pt}+ \frac{1}{16} \left[ 
         p^2 \left( -\frac{3}{8} + \frac{1}{2}\log(m^2)\right)
        -m^2 \left( \zeta (2) + \frac{7}{2} -3\log(m^2) + \log^2(m^2)
              \right) \right] \nonumber \\
        &&+ {\cal{F}}_0(m^2,0,0,p^2) + ... \Biggr\} \ . 
\labbel{205} \end{eqnarray}
In terms of \( {\cal{F}}_0(m^2,0,0,p^2) \) \Eq{204} then reads 
 \begin{equation}
  p^2(p^2+m^2) \frac{\partial^2}{\partial (p^2)^2} {\cal{F}}_0(m^2,0,0,p^2) 
    + 2 m^2  \frac{\partial}{\partial p^2} {\cal{F}}_0(m^2,0,0,p^2) 
  = \frac{1}{32} p^2 \ , 
\labbel{208} \end{equation}
which is in fact a first order differential equation in 
  \( \frac{\partial}{\partial p^2} {\cal{F}}_0(m^2,0,0,p^2) \). 
The values of \( {\cal{F}}_0(m^2,0,0,p^2) \) and its derivative at 
\( p^2=0 \) are by construction 
\begin{eqnarray} 
  {\cal{F}}_0(m^2,0,0,0) &=& 0  \ , \nonumber \\ 
  \frac{\partial}{\partial p^2}{\cal{F}}_0(m^2,0,0,0) &=& 0 \ . 
\labbel{211} \end{eqnarray} 
With the substitution
\begin{equation}
\frac{\partial}{\partial p^2} {\cal{F}}_0(m^2,0,0,p^2) 
        = \frac{(p^2+m^2)^2}{(p^2)^2} G(m^2,p^2)  \ \ , 
\labbel{212} \end{equation}
\Eq{208} becomes 
\begin{equation}
 \frac{\partial}{\partial p^2}G(m^2,p^2) =  
 \frac{1}{32}\frac{(p^2)^2}{(p^2+m^2)^3} \ \ .
\labbel{213} \end{equation} 
The integration of \Eq{213} is trivial; the result in terms of 
\( {\cal{F}}_0(m^2,0,0,p^2)  \), accounting for the second of the 
initial condition of \Eq{211}, reads 
\begin{equation}
\frac{\partial}{\partial p^2} {\cal{F}}_0(m^2,0,0,p^2)  =  
\frac{1}{32} \left[ \frac{(p^2+m^2)^2}{(p^2)^2} \log(1+\frac{p^2}{m^2}) 
 - \frac{m^2}{p^2} - \frac{3}{2} \right]  \ \ . 
\labbel{214} \end{equation}
One more integration and the first initial condition of \Eq{211} 
give 
\begin{equation}
 {\cal{F}}_0(m^2,0,0,p^2) = \frac{1}{32} \Biggl\{ -2 m^2 
      \hbox{Li}_2\left( -\frac{p^2}{m^2} \right) 
      +\left( p^2-\frac{m^4}{p^2} \right) 
      \log\left(1+\frac{p^2}{m^2}\right) 
      -\frac{5}{2}p^2 +m^2 \Biggr\} \ .  
\labbel{202} \end{equation}         \par 
The same result can be obtained, by means of the doubly subtracted 
dispersion relations \Eq{40} for the specific values of the masses, 
\begin{equation}
 {\cal{F}}_0(m^2,0,0,p^2) = \left(p^2\right)^2 \int\limits_{m^2}^{\infty}
 \frac{du}{u^2 (u+p^2)} \ \ \frac{1}{\pi} \ {\mathrm{Im}} F_0(4,m^2,0,0,-u)\ . 
\labbel{201} \end{equation}
The Cutkosky-Veltman rule gives for the imaginary part 
\begin{eqnarray}
 \frac{1}{\pi} \ {\mathrm{Im}} F_0(4,m^2,0,0,-u) &=& \frac{1}{16} 
 \int\limits_{m^2}^{u} db \ \ \frac{u-b}{u} \ \ \frac{b-m^2}{b} \nonumber \\
 &=& \frac{1}{16} \left\{ 
     \frac{u^2-m^4}{2u} - m^2\log{\frac{u}{m^2}} \right\}  \ ; 
\labbel{201a} \end{eqnarray}
substituting in \Eq{201} and carrying out a last integration \Eq{202} is 
recovered. 
\section{Large \(p^2\) expansion.}           \par 
By using the differential equations (\ref{5},\ref{7}) it is also possible 
to study the asymptotic expansion of the 4 functions \( \F{i} \) for large 
values of \( p^2 \). 
As \( p^2 = \infty \) is a singular point of the equations, we look for a 
tentative asymptotic expansion of \( F_0 \) in the form
\begin{equation} 
 F_0^{(\alpha)}(n,m_1^2,m_2^2,m_3^2,p^2) = 
(p^2)^{\alpha}  \sum \limits_{k=0}^{\infty} 
   \Phi_k^{(\alpha)}(n,m_1^2,m_2^2,m_3^2) \frac{1}{(p^2)^k}        \ .
\labbel{100} \end{equation}
Due to the linearity of the equations, the actual solution is built by 
summing a suitable linear combination of all the possible expansions. 
The corresponding expansions of the other 3 functions 
\( F_i^{(\alpha)}(n,m_1^2,m_2^2,m_3^2,p^2) \), \( i=1,2,3 \), 
can be obtained through \Eq{3}. \par 
By inserting the expansion \Eq{100} into \Eq{5} and \Eq{7}), 
one obtains a set of 4 equations, which must be satisfied, for each separate 
power of \( p^2 \), by the coefficients of the expansion. \\ 
A first solution corresponds to \( \alpha = - 1 \); we write it as 
\begin{equation} 
 \left\{ \begin{array}{l} 
 \alpha = - 1 \equiv r \ , \\ 
\frac{1}{p^2} \sum \limits_{k=0}^{\infty} 
\phi^{(r)}_k(n,m_1^2,m_2^2,m_3^2) \frac{1}{(p^2)^k}    \ . 
\end{array}\right. 
\labbel{100a} \end{equation} 
It corresponds to the socalled regular solution, hence the superscript $(r)$ 
in $\phi^{(r)}_k$ (we write $\phi^{(r)}_k$ rather than $\Phi^{(r)}_k$ 
for uniformity with later use). Its coefficients are completely determined 
by the non-homogeneous part of the equations; correspondingly, all the 
terms behave for large \( p^2 \) as strictly integer powers of \( 1/p^2 \), 
which is exactly the behaviour of the various terms occurring in the 
asymptotic expansion of the non-homogeneous part. 
The first coefficient of the expansion is 
\begin{equation} 
  \phi^{(r)}_0(n,m_1^2,m_2^2,m_3^2) = T(n,m_1^2) T(n,m_2^2)
  + T(n,m_1^2) T(n,m_3^2) + T(n,m_2^2) T(n,m_3^2) \ ; 
\labbel{112} \end{equation} 
all the other coefficients $\phi^{(r)}_k$ can be similarly obtained, when 
needed, by following the procedure just outlined. \\ 
Other solutions correspond to different values of \( \alpha \); if 
\( \alpha \) is not equal to \( -1 \), the leading coefficient must 
satisfy the homogeneous system of equations 
\begin{eqnarray} 
 &&\sum \limits_{i=1}^{3} m_i^2 \frac{\partial}{\partial m_i^2} 
    \Phi_0^{(\alpha)}(n,m_1^2,m_2^2,m_3^2)
    + \left(\alpha -n + 3 \right) \Phi_0^{(\alpha)}(n,m_1^2,m_2^2,m_3^2) = 0
        \nonumber \\ 
 && \left[ \alpha - \frac{1}{2} \left(n-4\right)
   \right]\frac{\partial}{\partial m_1^2} 
    \Phi_0^{(\alpha)}(n,m_1^2,m_2^2,m_3^2) = 0 \nonumber \\
 && \left[ \alpha - \frac{1}{2} \left(n-4\right)
   \right]\frac{\partial}{\partial m_2^2} 
    \Phi_0^{(\alpha)}(n,m_1^2,m_2^2,m_3^2) = 0 \nonumber \\
  &&\left[ \alpha - \frac{1}{2} \left(n-4\right)
   \right]\frac{\partial}{\partial m_3^2} 
    \Phi_0^{(\alpha)}(n,m_1^2,m_2^2,m_3^2) = 0 \ . 
\labbel{101} \end{eqnarray}
There are two solutions of the system (\ref{101}), corresponding to the 
values \( \alpha = (n-3) \) and \( \alpha = (n-4)/2 \). Correspondingly, 
Eq.s(\ref{101}) become 
\begin{equation} 
 \left\{ \begin{array}{l} 
 \alpha = n-3 \equiv s_1 \ , \\ 
 \frac{\partial}{\partial m_i^2} \Phi^{(s_1)}_0(n,m_1^2,m_2^2,m_3^2) = 0
    \hskip 5truemm (i = 1,2,3) \ , 
\end{array}\right. 
\labbel{103} \end{equation} 
and 
\begin{equation} 
 \left\{ \begin{array}{l} 
 \alpha = \frac{1}{2}\left(n-4\right) \equiv s_2 \ , \\ 
 \sum \limits_{i=1}^{3} m_i^2 \frac{\partial}{\partial m_i^2} 
    \Phi^{(s_2)}_0(n,m_1^2,m_2^2,m_3^2)
    - \frac{n-2}{2} \Phi^{(s_2)}_0(n,m_1^2,m_2^2,m_3^2) = 0 \ ; 
\end{array}\right. 
\labbel{102} \end{equation} 
the above values of \( \alpha \) are also called ``fractionary", 
and the corresponding solutions singular, hence the superscripts \( (s_1), 
(s_2) \), as opposed to the superscript $(r)$ for the ``regular" solution
corresponding to the integer value \( \alpha=-1 \) of \Eq{100a}. 
Due to the homogeneity of the equations for the coefficients of the 
``fractional" powers, the two leading coefficients 
\( \Phi^{(s_i)}_0(n,m_1^2,m_2^2,m_3^2) \) cannot be determined by the 
equations (\ref{103}) and (\ref{102}) alone; nevertheless, the equations 
can still provide much valuable information. \par 
\Eq{103} shows indeed that \( \Phi^{(s_1)}_0(n,m_1^2,m_2^2,m_3^2) \) is 
independent of the masses 
\begin{equation} \Phi^{(s_1)}_0(n,m_1^2,m_2^2,m_3^2) = \Phi^{(s_1)}(n) \ , 
\labbel{103a} \end{equation} 
so that the still unknown function \( \Phi^{(s_1)}(n) \) depends only 
on \( n \). \par 
The case of \Eq{102} is slightly more complicated; by investigating the 
condition on the leading \( (p^2)^{(n-4)/2} \) term implied by \Eq{301} 
of the Appendix, whose {\it l.h.s. } is the second derivative of $\F{0}$ 
with respect to $m_1^2$ and $m_2^2$, one finds 
the relations 
\begin{equation} 
  \frac{\partial^2}{\partial m_i^2 \partial m_j^2}
      \Phi^{(s_2)}_0(n,m_1^2,m_2^2,m_3^2) = 0
    \hskip 5truemm (i,j = 1,2,3 ; i \not= j) \ .
\labbel{106} \end{equation} 
According to \Eq{106}, the solution of \Eq{102} can be written as 
\begin{eqnarray} 
  \Phi^{(s_2)}_0(n,m_1^2,m_2^2,m_3^2) && = \Phi^{(s_2)}(n) 
                      \phi^{(s_2)}_0(n,m_1^2,m_2^2,m_3^2) \ , \nonumber\\ 
  \phi^{(s_2)}_0(n,m_1^2,m_2^2,m_3^2) && = 
                   m_1^{n-2} + m_2^{n-2} + m_3^{n-2} \ , 
\labbel{107} \end{eqnarray} 
where \( \Phi^{(s_2)}(n) \) is a function of \( n\) only. \par 
For each ``fractional" power solution, all the next-to-leading terms of the 
expansion are determined in terms of the leading terms, by solving recursively 
the systems of linear equations; it is therefore convenient to factorize 
the (as yet arbitrary) function $\Phi^{(s_i)}(n)$, defining  
\begin{equation} 
  \Phi^{(s_i)}_k(n,m_1^2,m_2^2,m_3^2) = \Phi^{(s_i)}(n) 
                      \phi^{(s_i)}_k(n,m_1^2,m_2^2,m_3^2) \ . 
\labbel{107a} \end{equation} 
An explicit calculation then gives for the first few terms (which we will use 
later) 
\begin{eqnarray}
 &&{\kern-20pt}\phi^{(s_1)}_1(n,m_1^2,m_2^2,m_3^2)  = \frac{(n-3)(3n-8)}{(n-4)}
  \left(m_1^2 + m_2^2 + m_3^2\right) \ ,         \nonumber\\
  &&{\kern-20pt}\phi^{(s_1)}_2(n,m_1^2,m_2^2,m_3^2) = \frac{(n-3)(3n-8)}{(n-6)}
  \Biggl[ - \frac{1}{(n-4)}B(m_1^2,m_2^2,m_3^2)
   + \R{2} \nonumber \\
   &&{\kern-20pt}- \frac{1}{2}B(m_1^2,m_2^2,m_3^2) 
    +\frac{3}{2}(n-4)\left(\R{2} + \frac{1}{2}B(m_1^2,m_2^2,m_3^2) \right)
    \Biggr] \ , 
\labbel{109} \end{eqnarray}
where 
\begin{equation} 
   B\left(m_1^2,m_2^2,m_3^2\right) = 4 
  \left( m_1^2 m_2^2 + m_1^2 m_3^2 + m_2^2 m_3^2 \right)
  \ , 
\labbel{110} \end{equation}
and 
\begin{eqnarray}
 &&{\kern-20pt}\phi^{(s_2)}_1(n,m_1^2,m_2^2,m_3^2) =
  \frac{n-3}{n} \Biggl[ -(3n-8)\left(m_1^2 + m_2^2 + m_3^2\right)
            \phi^{(s_2)}_0(n,m_1^2,m_2^2,m_3^2) \nonumber\\
 &&{\kern-20pt}+2(n-2)
   \Bigl( m_1^{n-2} (m_1^2+2m_2^2+2m_3^2)
        + m_2^{n-2} (2m_1^2+m_2^2+2m_3^2) \nonumber \\
   &&{\kern+40pt}  + m_3^{n-2} (2m_1^2+2m_2^2+m_3^2) \Bigr) \Biggr] \ . 
\labbel{111} \end{eqnarray}
Any subsequent term can be easily obtained, when needed, by the same procedure. 
\par 
As for arbitrary \( n \) the three solutions satisfy the equations 
independently of each other, the actual asymptotic expansion of \( \F{0} \) 
can be written as their sum 
\begin{eqnarray} 
  F_0^{(\infty)}(n,m_1^2,m_2^2,m_3^2,p^2) = 
   &&{\kern-10pt} p^2\ (p^2)^{n-4} \Phi^{(s_1)}(n) 
    \left( 1 + \sum \limits_{k=1}^{\infty} 
    \phi^{(s_1)}_k(n,m_1^2,m_2^2,m_3^2) \frac{1}{(p^2)^k} \right) \nonumber \\
   &&{\kern-10pt} +  (p^2)^{\frac{1}{2}(n-4)} \Phi^{(s_2)}(n) 
     \sum \limits_{k=0}^{\infty} 
   \phi^{(s_2)}_k(n,m_1^2,m_2^2,m_3^2) \frac{1}{(p^2)^k} \nonumber \\
   &&{\kern-10pt} + \frac{1}{p^2} \sum 
   \limits_{k=0}^{\infty} \phi^{(r)}_k(n,m_1^2,m_2^2,m_3^2) 
   \frac{1}{(p^2)^k} \ ; 
\labbel{108} \end{eqnarray} 
the above expression depends on the two functions of \( n \), 
\( \Phi^{(s_1)}(n) \) and \( \Phi^{(s_2)}(n) \), which are as yet 
undetermined. To fix them, we must supply more information on \( \F{0} \). 
As the \( \F{i}, i=0,1,2,3, \) were studied in the \( n\to4 \) limit, 
\Eq{39a}, we expand also \Eq{108} around \( n=4 \), writing in particular 
\( \Phi^{(s_1)}(n) \) and \( \Phi^{(s_2)}(n) \) as 
\begin{equation}
 \Phi^{(s_i)}(n) = C^2(n) 
   \sum \limits_{l=-2}^{\infty}(n-4)^k \Phi^{(s_i,l)} \ \ (i=1,2)
  \ ; 
\labbel{113} \end{equation}  
as always \( C(n) \) is not expanded, appearing as an overall factor. \\
By comparison with the coefficients of the singularities in $(n-4)$, 
which are exactly known and given in \Eq{39b}, the following values are found
\begin{eqnarray} 
  \Phi^{(s_1,-2)} = 0 \ , && \Phi^{(s_2,-2)} = -\frac{1}{4} \nonumber \\
  \Phi^{(s_1,-1)} = \frac{1}{32} \ , && 
  \Phi^{(s_2,-1)} = - 4\Phi^{(s_1,0)} - \frac{1}{32} \ .
\labbel{114} \end{eqnarray}  
Unfortunately some additional information has still to be supplied, 
besides that coming from the singularities in \( (n-4) \), for fixing 
the other as yet unknown constants \( \Phi^{(s_1,0)} \), 
\( \Phi^{(s_2,0)} \) and \( \Phi^{(s_1,1)} \); the last quantity is 
the coefficient of the term in \( (n-4) \) in \Eq{113}, which is also 
needed due to the presence of the \( 1/(n-4) \) factor in \Eq{109}.  \par 
As those constants are independent of the masses, they can be determined 
through the comparison with the asymptotic expansion of a simpler case, 
namely \( F_0(n,m^2,0,0,p^2) \), discussed in the previous Section, whose 
explicit analytic value is given in \Eq{202}. 
Its large \( p^2 \) expansion reads 
\begin{eqnarray}
 F_0^{(\infty)}(n,m^2,0,0,p^2) &&{\kern-15pt}= C^2(n) \Biggl\{ 
  - \frac{1}{(n-4)^2} \frac{m^2}{8} 
  + \frac{1}{(n-4)} \frac{1}{32}\left(p^2 
        + 6m^2 -4m^2 \log(m^2)\right) \nonumber \\
  &&{\kern-40pt}+ \frac{1}{16} \Biggl[ 
         p^2 \left( -\frac{13}{8} + \frac{1}{2}\log(p^2)\right)
     +\frac{m^2}{2} \log^2(p^2) - m^2 \log(m^2)\log(p^2)
 \nonumber \\
 &&{\kern-2pt} +m^2\left(-\frac{1}{2} \log^2(m^2) + 3 \log(m^2)
      -\frac{5}{2} 
      \right)\nonumber \\
 &&{\kern-2pt} 
    -\frac{m^4}{2p^2}\left( \log(p^2) - \log(m^2) + \frac{5}{2}\right)
    \Biggr] + ... \Biggr\} \ \ . 
\labbel{203} \end{eqnarray} 
Let us recall that, when using \Eq{108} and \Eq{113}, the logarithmic factors 
are generated in the \( n\to4 \) limit through expansions of the kind 
\[ \frac{1}{(n-4)^2} (p^2)^{n-4} = \frac{1}{(n-4)^2} 
       \left( 1 + (n-4)\log(p^2) + \frac{1}{2}(n-4)^2 \log^2(p^2) 
       + \cdots \right) \ . \] 
By comparing \Eq{203} and the \( n\to4 \) limit of \Eq{108}, when two 
masses are taken equal to zero, \Eq{114} is completed into 
\begin{eqnarray} 
  \Phi^{(s_1,-2)} &&{\kern-20pt}= 0 \ \ , \ \ \ \ \ \ \ \ \ \ \ \
  \Phi^{(s_2,-2)} = -\frac{1}{4}    \nonumber \\
  \Phi^{(s_1,-1)} &&{\kern-20pt}= \frac{1}{32}    \ \ , \ \ \ \ \ 
      \ \ \ \ \ 
  \Phi^{(s_2,-1)} = \frac{3}{8}     \nonumber \\
  \Phi^{(s_1,0)} &&{\kern-20pt}= -\frac{13}{128} \ \ , \ \ \ \ \ \ 
  \Phi^{(s_2,0)} + 4\Phi^{(s_1,1)}  = \frac{59}{128}
  \ .
\labbel{114.114} \end{eqnarray}  
It is apparent from the above equations that only a particular combination 
of \( \Phi^{(s_2,0)} \) and \( \Phi^{(s_1,1)} \) is determined by 
the comparison, but only that combination is needed for expressing 
\( \F{0} \) up to the constant term in \( (n-4) \) ( {\it i.e. } the 
separate knowledge of the two terms is not needed up to the degree of 
expansion in \( (n-4) \) and \( p^2 \), considered here). 
\par 
Summing up, for arbitrary masses and in the \( n \to 4 \) limit, 
up to terms \( 1/p^2 \) included, the asymptotic expansion \Eq{108} 
can be written as 
\begin{eqnarray} 
 F_0^{(\infty)} (m_1^2,m_2^2,m_3^2,p^2) &=& C^2(n) \Biggl\{ \frac{1}{(n-4)^2} 
           F_0^{(\infty,-2)}(m_1^2,m_2^2,m_3^2,p^2) \nonumber\\ 
       &+& \frac{1}{(n-4)} F_0^{(\infty,-1)} (m_1^2,m_2^2,m_3^2,p^2) 
           \nonumber\\ 
       &+& F_0^{(\infty,0)} (m_1^2,m_2^2,m_3^2,p^2) + ... \Biggr\} \ ,
\labbel{115} \end{eqnarray}  
with 
\begin{equation} 
 F_0^{(\infty,-2)} (m_1^2,m_2^2,m_3^2,p^2) = - \frac{1}{8}
             \left(m_1^2 + m_2^2 + m_3^2 \right) \ ,
\labbel{116} \end{equation}  
\begin{equation} 
 F_0^{(\infty,-1)} (m_1^2,m_2^2,m_3^2,p^2) = \frac{1}{16} 
     \left[ \frac{1}{2}p^2 
  + \sum \limits_{i=1}^{3}m_i^2 \left(3 - 2 \log(m_i^2)\right) \right] \ , 
\labbel{117} \end{equation}  
and 
\begin{eqnarray} 
 F_0^{(\infty,0)} (m_1^2,m_2^2,m_3^2,p^2) = 
   &&{\kern-20pt} p^2 \left[ f_{1,1} \log(p^2) + f_{1,0} \right] \nonumber \\
 + &&{\kern-20pt} f_{0,2} \log^2(p^2) + f_{0,1} \log(p^2) + f_{0,0} \nonumber \\
 + &&{\kern-20pt} \frac{1}{p^2} \left[ f_{-1,2} \log^2(p^2)   
  + f_{-1,1} \log(p^2) + f_{-1,0} \right] + \  \ ... \ ,
\labbel{118} \end{eqnarray}  
where 
\begin{eqnarray} 
 &&{\kern-20pt} f_{1,1} = \frac{1}{32} \ \ , \ \ \ f_{1,0} = -\frac{13}{128} 
   \ \ , \ \ \
   f_{0,2} = \frac{1}{32}\sum\limits_{i=1}^{3}m_i^2 \ \ , \ \ \
   f_{0,1} = -\frac{1}{16}\sum\limits_{i=1}^{3}m_i^2\log(m_i^2) \  , 
   \nonumber \\
   &&{\kern-20pt}     f_{0,0} =    - \frac{1}{32}
  \sum\limits_{i=1}^{3}m_i^2\left(5 -6\log(m_i^2)+\log^2(m_i^2)\right)
   \ \ , \ \ \
   f_{-1,2} = \frac{1}{64}B(m_1^2,m_2^2,m_3^2)  \nonumber \\
   &&{\kern-20pt}f_{-1,1} = \frac{1}{64} 
                            \Bigl(B(m_1^2,m_2^2,m_3^2) -2 \R{2}  
    - 4(m_2^2+m_3^2)m_1^2 \log(m_1^2)   \nonumber \\
     &&{\kern+60pt}  
    - 4(m_1^2+m_3^2)m_2^2 \log(m_2^2)
    - 4(m_1^2+m_2^2)m_3^2 \log(m_3^2) \Bigr) \nonumber \\
&&{\kern-20pt} f_{-1,0} = \frac{1}{128} \Bigl(-3 B(m_1^2,m_2^2,m_3^2) -10 \R{2} 
   +4 ( m_1^2 - 2m_2^2 - 2m_3^2) m_1^2\log(m_1^2)   \nonumber \\
      &&{\kern+60pt}+4 ( m_2^2 - 2m_1^2 - 2m_3^2) m_2^2\log(m_2^2) 
     +4 ( m_3^2 - 2m_1^2 - 2m_2^2) m_3^2\log(m_3^2)  \nonumber \\   
     &&{\kern+60pt}+8 m_1^2 m_2^2\log(m_1^2) \log(m_2^2)
     +8 m_1^2 m_3^2\log(m_1^2) \log(m_3^2) \nonumber \\
     &&{\kern+60pt}+8 m_2^2 m_3^2\log(m_2^2) \log(m_3^2)
       \Bigr) 
     \ .
\labbel{119} \end{eqnarray}  
The coefficients of \( 1/(n-4)^2 \) and \( 1/(n-4) \), \Eq{116} and \Eq{117}, 
are of course identical to the exact ones in \Eq{39b}), as their expansions 
in \(1/p^2\) contain only a linear and a constant term. The results given 
in Eq.s(\ref{118},\ref{119}) agree with the first terms of the corresponding 
expansion given in \cita{Berends}. 
\par 
Higher order terms, when needed, can be added with little effort by 
following the above discussed procedure. 
\par 
\vskip 1 truecm  
\noindent{\bf {\large Acknowledgments.} }\par
One of us (HC) is grateful to the Bologna Section of INFN and to the 
Department of Physics of Bologna University for support and kind 
hospitality.
\bigskip

\noindent\ \appendix{\bf {\large Appendix.} }           \par 
We list here a few ready-to-use formulae, obtained by solving 
the integration by part identities for the amplitudes of the sunrise 
self-mass graph (see also \cita{Tarasov}). The notation for the amplitudes 
is given in \Eq{1}. Further relations can be found by permutation of masses.
\begin{eqnarray} 
 && 
 D(m_1^2,m_2^2,m_3^2,p^2) A(n,m_1^2,m_2^2,m_3^2,p^2,-3,-1,-1,0,0) =
   \nonumber \\
 &&{\kern+50pt} -\frac{n-4}{4m_1^2}D(m_1^2,m_2^2,m_3^2,p^2) \F{1} \nonumber \\
 &&{\kern+50pt} +(n-3)(p^2+m_1^2) 
    \Bigl[ 4(m_1^2 m_2^2 + m_1^2 m_3^2 - m_2^2 m_3^2) \nonumber \\
   &&{\kern+150pt} - (p^2+m_1^2+m_2^2+m_3^2)^2 \Bigr] \F{1} \nonumber \\
 &&{\kern-30pt} +(n-3)(p^2+m_2^2)\frac{m_2^2}{m_1^2}
   \Bigl[ 4(m_1^2 m_2^2 - m_1^2 m_3^2 + m_2^2 m_3^2)\nonumber \\
    &&{\kern+150pt}- (p^2+m_1^2+m_2^2+m_3^2)^2 \Bigr] \F{2} \nonumber \\
 &&{\kern-30pt} +(n-3)(p^2+m_3^2)\frac{m_3^2}{m_1^2}
   \Bigl[ 4(-m_1^2 m_2^2 + m_1^2 m_3^2 + m_2^2 m_3^2) \nonumber \\
   &&{\kern+150pt} - (p^2+m_1^2+m_2^2+m_3^2)^2 \Bigr] \F{3} \nonumber \\
 &&{\kern-30pt} +(n-3)(3n-8)\frac{1}{m_1^2}
   \Bigl[  -4m_1^2m_2^2m_3^2 
  +(m_1^2 m_2^2 + m_1^2 m_3^2 + m_2^2 m_3^2)(p^2+m_1^2+m_2^2+m_3^2) \nonumber \\
   &&{\kern+140pt}
      -\frac{1}{4}(p^2+m_1^2+m_2^2+m_3^2)^3 \Bigr] \F{0} \nonumber \\
 &&{\kern-30pt} +\frac{(n-2)^2}{m_1^2}\left[ (m_3^2-m_1^2)(m_3^2-m_2^2)
     - \frac{1}{4} (p^2+m_1^2+m_2^2-m_3^2)^2 \right] 
    T(n,m_1^2) T(n,m_2^2) \nonumber \\
 &&{\kern-30pt} +\frac{(n-2)^2}{m_1^2}\left[ (m_2^2-m_1^2)(m_2^2-m_3^2)
     - \frac{1}{4} (p^2+m_1^2-m_2^2+m_3^2)^2 \right] 
    T(n,m_1^2) T(n,m_3^2) \nonumber \\
   &&{\kern-30pt}+ \frac{(n-2)^2}{m_1^2}\left[ (m_1^2-m_2^2)(m_1^2-m_3^2)
     - \frac{1}{4} (p^2-m_1^2+m_2^2+m_3^2)^2 \right] 
    T(n,m_2^2) T(n,m_3^2)  \ \ ,
 \labbel{300} \end{eqnarray}

\vfill
\eject

\begin{eqnarray}
  &&{\kern-30pt}D(m_1^2,m_2^2,m_3^2,p^2) A(n,m_1^2,m_2^2,m_3^2,p^2,-1,-2,-2,0,0) =
   \nonumber \\
  &&{\kern+100pt}8 (n-3) m_1^2 (p^2+m_1^2)(p^2-m_1^2+m_2^2+m_3^2) \F{1} 
                                                           \nonumber\\
  &&{\kern-30pt} -(n-3)\Bigl[(p^2-m_1^2+m_2^2+m_3^2)^3 
              -4(m_2^2-m_1^2)(p^2-m_1^2+m_2^2+m_3^2)^2 \nonumber \\
     &&{\kern+50pt} - 4(m_3^2-m_1^2)(m_1^2+m_2^2)(p^2-m_1^2+m_2^2+m_3^2) \nonumber \\
     &&{\kern+160pt} +16m_2^2(m_3^2-m_1^2)(m_2^2-m_1^2) \Bigr]\F{2} \nonumber \\
  &&{\kern-30pt} -(n-3)\Bigl[(p^2-m_1^2+m_2^2+m_3^2)^3 
              -4(m_3^2-m_1^2)(p^2-m_1^2+m_2^2+m_3^2)^2 \nonumber \\ 
      &&{\kern+50pt}- 4(m_2^2-m_1^2)(m_1^2+m_3^2)(p^2-m_1^2+m_2^2+m_3^2) \nonumber \\
      &&{\kern+160pt}+16m_3^2(m_3^2-m_1^2)(m_2^2-m_1^2) \Bigr] \F{3} \nonumber \\
  &&{\kern-30pt} +(n-3)(3n-8)\Bigl[ (p^2-m_1^2+m_2^2+m_3^2)^2 -4(m_3^2 -m_1^2)(m_2^2-m_1^2)
                \Bigr] \F{0} \nonumber \\
  &&{\kern-30pt} -\frac{(n-2)^2}{2m_2^2}\Bigl[(p^2+m_1^2-m_2^2+m_3^2)^2 
            +4(m_3^2-m_2^2)(m_2^2-m_1^2) \Bigr] 
              T(n,m_1^2) T(n,m_2^2) \nonumber \\
  &&{\kern-30pt} -\frac{(n-2)^2}{2m_3^2}\Bigl[(p^2+m_1^2+m_2^2-m_3^2)^2 
            +4(m_2^2-m_3^2)(m_3^2-m_1^2) \Bigr] 
              T(n,m_1^2) T(n,m_3^2) \nonumber \\
 &&{\kern-30pt} +\frac{(n-2)^2}{4m_2^2 m_3^2}\Bigl[(p^2+m_1^2+m_2^2+m_3^2)^3
       - 4(m_1^2m_2^2 + m_2^2m_3^2 + m_1^2m_3^2)(p^2+m_1^2+m_2^2+m_3^2)
     \nonumber \\
       &&{\kern+200pt}+16 m_1^2 m_2^2 m_3^2 \Bigr] T(n,m_2^2) T(n,m_3^2) \ \ .
\labbel{301} \end{eqnarray}
\vfill \eject 
\def\NP{{\sl Nuc. Phys.}\ } 
\def\PL{{\sl Phys. Lett.}\ } 
\def\PR{{\sl Phys. Rev.}\ } 
\def\PRL{{\sl Phys. Rev. Lett.}\ } 

\end{document}